\documentclass{article}

\usepackage{arxiv}

\usepackage[utf8]{inputenc} 
\usepackage[T1]{fontenc}    
\usepackage{hyperref}       
\usepackage{url}            
\usepackage{booktabs}       
\usepackage{amsfonts}       
\usepackage{nicefrac}       
\usepackage{microtype}      
\usepackage{lipsum}		
\usepackage{graphicx}
\usepackage{natbib}
\usepackage{doi}
\usepackage{amsmath}
\usepackage{svg}
\usepackage[affil-it]{authblk}

\title{Gradients of unitary optical neural networks using parameter-shift rule}

\author[ ,1,2]{Jinzhe Jiang \thanks{jiangjinzhe@ieisystem.com}  
}
\author[1,2]{Yaqian Zhao}
\author[1,2]{Xin Zhang}
\author[1,2]{Chen Li}
\author[1,2]{Yunlong Yu}
\author[1,2]{Hailing, Liu}
\affil[1]{IEIT SYSTEMS Co., Ltd, No.801, Caoshan Lingnan Road, High-tech Zone, Jinan, Shandong, 250101, China }
\affil[2]{IEIT SYSTEMS (Beijing) Co., Ltd, No.2, Shangdi Information Road, Haidian District, Beijing, 100095, China}

\renewcommand{\shorttitle}{\textit{arXiv} Template}

\hypersetup{
pdftitle={A template for the arxiv style},
pdfsubject={q-bio.NC, q-bio.QM},
pdfauthor={David S.~Hippocampus, Elias D.~Striatum},
pdfkeywords={First keyword, Second keyword, More},
}

\begin{document}
\maketitle

\begin{abstract}
This paper explores the application of the parameter-shift rule (PSR) for computing gradients in unitary optical neural networks (UONNs). While backpropagation has been fundamental to training conventional neural networks, its implementation in optical neural networks faces significant challenges due to the physical constraints of optical systems. We demonstrate how PSR, which calculates gradients by evaluating functions at shifted parameter values, can be effectively adapted for training UONNs constructed from Mach-Zehnder interferometer meshes. The method leverages the inherent Fourier series nature of optical interference in these systems to compute exact analytical gradients directly from hardware measurements. This approach offers a promising alternative to traditional in silico training methods and circumvents the limitations of both finite difference approximations and all-optical backpropagation implementations. We present the theoretical framework and practical methodology for applying PSR to optimize phase parameters in optical neural networks, potentially advancing the development of efficient hardware-based training strategies for optical computing systems.
\end{abstract}


\section{Introduction}
The efficacy of contemporary artificial neural networks (ANNs) and the deep learning models they underpin is critically dependent on sophisticated training methodologies, with backpropagation standing as the foundational algorithm. Introduced as a key method for training these networks, backpropagation enables learning from data by systematically minimizing the discrepancy between network predictions and actual outcomes. The formalization of backpropagation for multi-layer networks by Rumelhart, Hinton, and Williams in 1986 was a watershed moment, demonstrating its capacity to learn "interesting" distribution representations and rendering the training of deep, complex networks computationally feasible \citep{rumelhart_1986_learning}.Prior to this, adjusting the millions of parameters often found in deep networks was a slow and arduous task, but backpropagation's efficient gradient computation transformed this landscape. The algorithm's power stems directly from the mathematical elegance of the chain rule, which allows for a methodical, layer-by-layer apportionment of error, making the complex problem of credit assignment in deep architectures tractable \citep{li_2024_comprehensive}. Its profound impact is evidenced by its widespread application across diverse domains, including image recognition and natural language processing, and it is fair to assert that the deep learning revolution is inextricably linked to the efficiency and scalability afforded by backpropagation.

In the quest for computational paradigms that can transcend the limitations of conventional electronics, optical computing have emerged as a compelling alternative, promising ultra-high processing speeds, diminished latency, extensive parallelism, and superior energy efficiency \citep{wetzstein_2020_inference}. And Optical Neural Networks (ONN) leverage light for computation, offering a physically distinct platform for artificial intelligence \citep{saadabadhaniehmasoudian_2025_physicsconstrained, apostolostsakyridis_2024_photonic, khonina_2024_exploring, wei_2024_spatially}. However, the translation of training algorithms, particularly backpropagation, to the optical domain is fraught with substantial challenges \citep{matuszewski_2024_role, fu_2025_symbiotic}. A primary impediment is the inherent requirement of backpropagation to perform calculations in the reverse order of the inference signal flow, a task that proves difficult to implement directly on analog physical systems like ONNs \citep{james_2025_training}. Consequently, ONNs are often trained in silico, where network parameters are optimized on a digital simulator and subsequently transferred to the optical hardware. Early explorations into optical backpropagation, such as the work by Hughes et al., made valuable contributions \citep{hughes_2018_training}.  Similarly, other pioneering ONN demonstrations frequently relied on in silico backpropagation for training the parameters of their coherent nanophotonic circuits, sometimes exploring alternative on-chip methods like finite differences for gradient estimation, like those by \citep{shen_2017_deep}. There have catalyzed a broader search for novel training methodologies better suited to the unique characteristics of optical hardware \citep{lu_2023_efficient}.

An alternative pathway for gradient computation is the parameter-shift rule (PSR). This technique allows for the analytical calculation of the partial derivative of a function by evaluating the function at specifically shifted values of the parameter under consideration \citep{markovich_2024_parameter}. Analogous concepts are emerging as relevant to classical linear optical circuits, particularly those whose output characteristics (like intensity or field amplitude) can be expressed as a finite Fourier series of their internal tunable parameters, such as the phase shifts in Mach-Zehnder Interferometers (MZIs). The output of MZI meshes, common building blocks for Unitary Optical Neural Networks (UONNs), inherently exhibits this trigonometric, and thus Fourier series, dependence on phase parameters due to wave interference \citep{facelli_2024_exact}. This shared mathematical structure-the decomposability of the output function into a finite sum of sinusoids-is the fundamental link that allows PSR-like methods to be effective in classical optical interferometers \citep{pastor_2021_arbitrary}. This Fourier property enables the analytical calculation of gradients with respect to these phase parameters by evaluating the circuit at shifted phase values, offering a route to exact gradients directly from hardware measurements. Such an approach could circumvent the approximation errors inherent in finite difference methods, especially in noisy environments, and the significant implementation hurdles of all-optical backpropagation. The extension of PSR to classical photonics thus represents a promising avenue, leveraging sophisticated mathematical tools from one domain to solve pressing challenges in another.

The paper commences by providing essential background on the parameter-shift rule, detailing its mathematical underpinnings, particularly for gate parameterizations relevant to optical phase shifters. It also offers an overview of unitary optical neural network architectures, emphasizing those constructed from Mach-Zehnder interferometer meshes. Subsequently, the methodology for training these UONNs is delineated. This involves adapting and applying the parameter-shift rule to compute the analytical gradients of a defined loss function with respect to the MZI phase parameters, thereby enabling gradient-based optimization. In the end followed by a concluding discussion.

\section{Background}
\label{sec:backgroud}

\subsection{Unitary Optical Neural Networks}
This paper investigates the application of the parameter-shift rule for training Unitary Optical Neural Networks (UONNs). UONNs represent a specialized class of ONNs wherein the network transformations are mathematically unitary, often realized using meshes of MZIs configured according to schemes like those by Clements or Reck. These architectures are noted for their potential robustness and suitability for physical implementation \citep{yezhang_2024_design, hamerly_2022_accurate, bell_2021_further}.

The standard unit of UONNs is Mach-Zehnder interferometer (MZI), consisting of two 50:50 beam splitters (BS) and two phase shifters (PS). The input state owns two paths, and the phase shifts are adjusted to control the output state. MZIs implement a 2 × 2 unitary transformation on the input state of the form: 

$$U_{MZI} = \frac{1}{2}\begin{bmatrix} e^{i\phi}(e^{i\theta}-1) & ie^{i\phi}(1+e^{i\theta}) \\ i(e^{i\theta}+1) & 1-e^{i\theta} \end{bmatrix}$$

\begin{figure}[htbp]
  \centering
  \includegraphics[trim={0cm 22cm 0cm 0cm}, clip, width=0.9\textwidth]{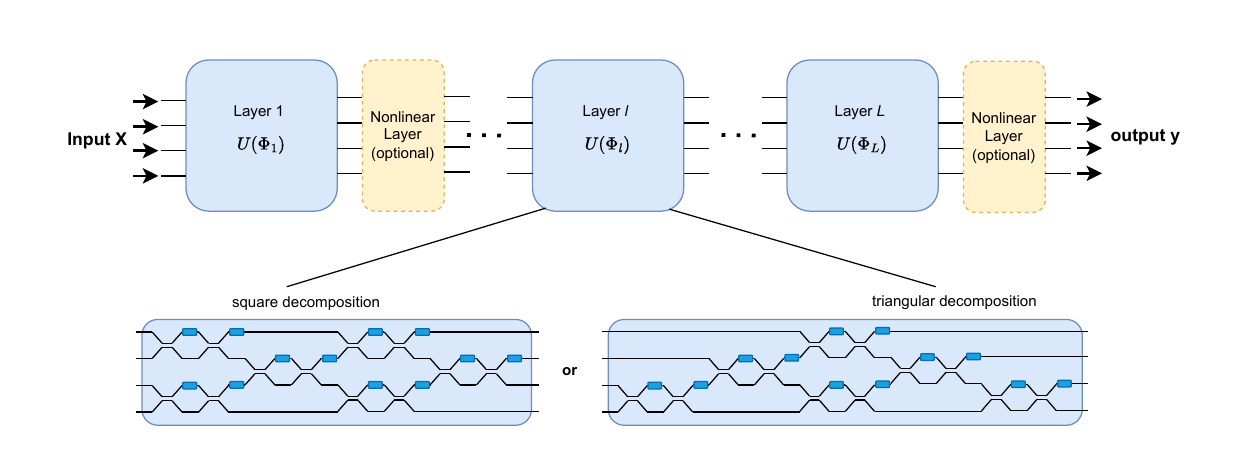}
  \caption{Schematic diagram of unitary optical neural network. The light blue rectangular boxes represent optical layers, and each optical layer completes a unitary matrix operation realized by MZI mesh, which can be decomposed with the triangular decomposition method or the square decomposition method. The light yellow rectangular boxes represent the nonlinear layers, which is optional based on the needs of the model.}
  \label{fig:1}
\end{figure}

By combining MZIs in a specific manner, high-dimensional data vectors can be processed. The triangular decomposition method by Reck et al. or the square decomposition method by Clements et al. can implement arbitrary unitary matrices \citep{reck_1994_experimental, clements_2016_an}. In general, these two mzi mesh structures are considered as the hidden layer  to form an unitary optical neural network structure, as shown in Fig.\ref{fig:1}.

\subsection{Parameter-Shift Rule}
The parameter-shift rule is a promising method in unitary evolution that allows for the efficient evaluation of the gradient of a function with respect to its parameters. The rule is based on the observation that the gradient of a function can be expressed as a linear combination of the function's values at shifted parameter values. Consider the unitary $U(\vec{\theta})$, which is formed by a product of unitary transformations $\prod_{j=1}^{l}U_j(\theta_j)$ acting on a state $\rho_{in}$ in Hilbert Space and we have the function $f(\vec{\theta})$ with an observable $O$: 

$$f(\vec{\theta})=Tr(OU_{l:1}\rho_{in}U_{1:l}^{\dagger})$$

Here $U_{j:k}=U_j...U_k$, then the gradient is calculated to be 
$$\frac{\partial f(\vec{\theta})}{\partial \theta_j} = Tr(OU_{l:j+1} \frac{\mathrm{d}U_j}{\mathrm{d}\theta_j} U_{j-1:1}\rho_{in}U_{1:l}^{\dagger}+OU_{l:1}\rho_{in}U_{1:j-1}^{\dagger}\frac{\mathrm{d}U_j^{\dagger}}{\mathrm{d}\theta_j}U_{j+1:l}^{\dagger})$$

\subsubsection{Pauli Generator}
Firstly, we consider the common situation in Hilbert Space,  $U_j(\theta)=e^{-i\theta_jP_j/2}$ is generated by a Pauli operator $P_j$. Then we have the derivative of each unitary transformation as:
$$\frac{\mathrm{d}U_j(\theta_j)}{\mathrm{d}\theta_j}=-i\frac{P_j}{2}U_j$$ 
$$\frac{\mathrm{d}U_j^{\dagger}(\theta_j)}{\theta_j}=i\frac{P_j}{2}U_j$$ 

The gradient becomes:

\begin{equation*}
\begin{split}
\frac{\partial f(\theta_j)}{\partial \theta_j} &= -\frac{i}{2}Tr(OU_{l:j+1}P_jU_{j:1}\rho_{in}U_{1:l}^{\dagger}-OU_{l:1}\rho_{in}U_{1:j}^{\dagger}P_jU_{j+1:l}^{\dagger})\\
&= -\frac{i}{2}Tr(OU_{l:j+1}P_jU_{j:1}\rho_{in}U_{1:j}^{\dagger}U_{j+1:l}^{\dagger}-OU_{l:j+1}U_{j:1}\rho_{in}U_{1:j}^{\dagger}P_jU_{j+1:l}^{\dagger})\\
&= -\frac{i}{2}Tr(OU_{l:j+1}P_j\rho_jU_{j+1:l}^{\dagger}-OU_{l:j+1}\rho_jP_jU_{j+1:l}^{\dagger})\\
&=-\frac{i}{2}Tr(OU_{l:j+1}(P_j\rho_j-\rho_jP_j)U_{j+1:l}^{\dagger})
\end{split}    
\end{equation*}

Here $\rho_j=U_{j:1}\rho_{in}U_{1:j}^{\dagger}$, and we can derive that:

$$P_j\rho_j-\rho_jP_j=i[U_j(\pi/2)\rho_j U_j^{\dagger}(\pi/2)-U_j(-\pi/2)\rho_j U_j^{\dagger}(-\pi/2)]$$

The gradient can be evaluated by:

\begin{equation*} 
\begin{split}
\frac{\partial f(\theta_j)}{\partial \theta_j} &=1/2Tr(OU_{l:j}U_j(\pi/2)\rho_{j}U_j^{\dagger}(\pi/2)U_{j:l}^{\dagger})-1/2Tr(OU_{l:j}U_j(-\pi/2)\rho_{j}U_j^{\dagger}(-\pi/2)U_{j:l}^{\dagger}) \\
&=1/2Tr(OU_{l:j}U_j(\pi/2)U_{j:1}\rho_{in}U_{1:j}^{\dagger}U_j^{\dagger}(\pi/2)U_{j:l}^{\dagger})-1/2Tr(OU_{l:j}U_j(-\pi/2)U_{j:1}\rho_{in}U_{1:j}^{\dagger}U_j^{\dagger}(-\pi/2)U_{j:l}^{\dagger})
\end{split} 
\end{equation*}

By inserting $\pm \frac{\pi}{2}$ rotation generated by $P_j$ and measuring the respective function values 
$$f(\theta_j + \frac{\pi}{2}) =Tr(OU_{l:j+1}U_j(\theta_j+\pi/2)U_{j-1:1}\rho_{in}U_{1:j-1}^{\dagger}U_{j}^{\dagger}(\theta_j+\pi/2)U_{j+1:l}^{\dagger})$$
$$f(\theta_j - \frac{\pi}{2}) =Tr(OU_{l:j+1}U_j(\theta_j-\pi/2)U_{j-1:1}\rho_{in}U_{1:j-1}^{\dagger}U_{j}^{\dagger}(\theta_j-\pi/2)U_{j+1:l}^{\dagger})$$
We can evaluate the exact gradient of the parameter $f(\theta_j)$:

$$\frac{\partial f(\theta_j)}{\partial \theta_j}=\frac{1}{2}[f(\theta_j + \frac{\pi}{2})-f(\theta_j - \frac{\pi}{2})]$$

\subsubsection{Hermitian Generator}
For more general case, the unitary transformation $U_j(\theta_j)=e^{-ia\theta_jH_j}$ is generated by a Hermitian operator $H_j$ and $a$ is a real constant. If the generator of the gate $H_j$ has only two unique eigenvalues $e_0$ and $e_1$, it can be converted to a unitary operator $G_j=\frac{a}{r}(H_j-s)$, where $r = \frac{a}{2}(e_1-e_0)$ and $s=\frac{1}{2}(e_1+e_0)$. And $U_j$ becomes $U_j=e^{-i\theta (rG_j + as)}$. 

For the same function $f(\vec{\theta})=Tr(OU_{l:1}\rho_{in}U_{1:l}^{\dagger})$, we substitute the shift parameter $\Delta \theta=\frac{\pi}{4r}$ into the equation $\frac{\mathrm{d}}{\mathrm{d}\theta}f(\theta)$ with the same derivation procedure of Pauli Genertor, then we can have the gradient of the parameter: 

$$\frac{\mathrm{d}}{\mathrm{d}\theta}f(\theta)=r[f(\theta+\frac{\pi}{4r})-f(\theta-\frac{\pi}{4r})]$$

\subsection{Opitcal Parameter-shift rule}
A MZI consists of four components: beamspliter $BS_1$, phaseshifter $PS_1$,  beamspliter $BS_2$ and phaseshifter $PS_2$. The transfer matrix of each component is:

$$U_{BS1},U_{BS2} =\frac{\sqrt{2}}{2}\left[\begin{matrix} 1 & i \\ i & 1 \end{matrix}\right]$$

$$U_{PS1} =\left[\begin{matrix} e^{i\theta_1} & 0 \\ 0 & 1 \end{matrix}\right], U_{PS2} =\left[\begin{matrix} e^{i\theta_2} & 0 \\ 0 & 1 \end{matrix}\right]$$

We can use electric field to describe input light vector $E_{input}=[E_{i1}, E_{i2}]^{\mathrm{T}}$ , and the output vector:
$$E_{output}=[E_{o1}, E_{o2}]^T=U_{PS2}U_{BS2}U_{PS1}U_{BS1}E_{input}=WE_{input}$$

\subsubsection{Real Output of ONN Layer}
\begin{figure}[htbp]
  \centering
  \includegraphics[trim={0cm 14cm 0cm 0cm}, clip, width=0.8\textwidth]{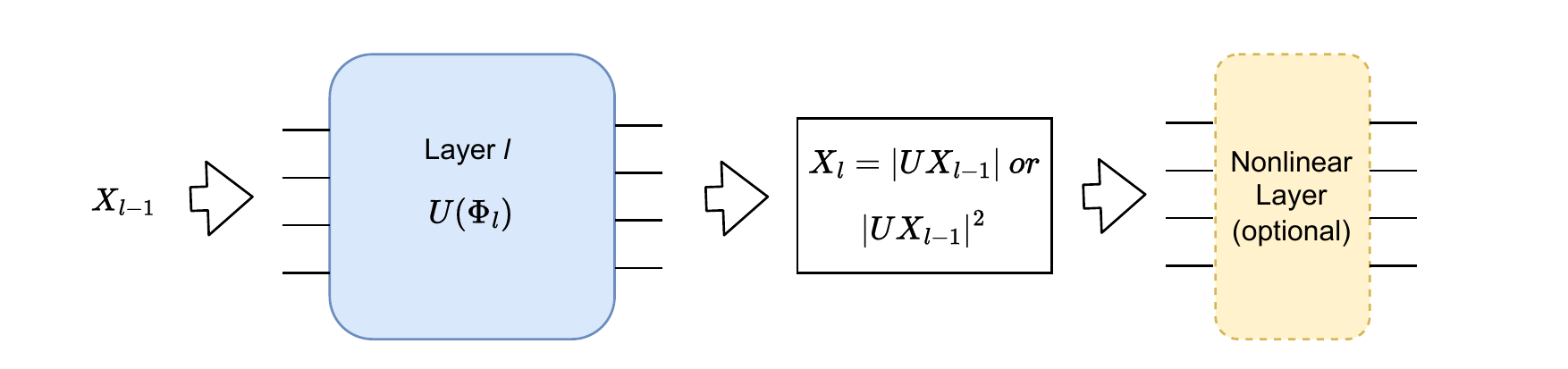}
  \caption{Schematic diagram of a optical layer with real-valued output. After detection, the intensity or amplitude of the light is used as the output of the layer and input to the next layer.}
  \label{fig:2}
\end{figure}

In most cases like Fig.\ref{fig:2}, we use photodiodes to measure the intensity of the output optical field, that is proportional to the square of the electric field amplitude, $|E_{o1}|^2$ and $|E_{o2}|^2$. The intensity of the output optical field can be expressed as:

$$|E_{o1}|^2=Tr(Z_1WE_{input}E_{input}^{\dagger}W^{\dagger})$$
$$|E_{o2}|^2=Tr(Z_2WE_{input}E_{input}^{\dagger}W^{\dagger})$$

where $Z_1$ and $Z_2$ are the diagonal matrices of the measurement basis, $Z_1=\left[\begin{matrix} 1 & 0 \\ 0 & 0 \end{matrix}\right]$ and $Z_2=\left[\begin{matrix} 0 & 0 \\ 0 & 1 \end{matrix}\right]$.

This is the same as the expression for $f(\theta)$ in the parameter-shift method. By further analyzing the expression of the parameterized phase shifter, we can derive that phaseshifter is generated by a Hermitian matrix $G$:

$$U_{PS}(\theta) =\left[\begin{matrix} e^{i\theta} & 0 \\ 0 & 1 \end{matrix}\right] = e^{-ia\theta G}$$

where $G=\left[\begin{matrix} 1 & 0 \\ 0 & 0 \end{matrix}\right]$ and has two eigenvalues $e_1=1, e_2=0$, and $a=-1$. 

According to the parameter shift rule of:
$$\frac{\mathrm{d}}{\mathrm{d}\theta}f(\theta)=r[f(\theta+\frac{\pi}{4r})-f(\theta-\frac{\pi}{4r})]$$
Where the shift constant is $r = \frac{a}{2}(e_1-e_0)$. We can obtain that $r=-\frac{1}{2}$ and:
$$\frac{\mathrm{d}}{\mathrm{d}\theta_1}|E_{o1}|^2=\frac{1}{2}[|E_{o1}(\theta_1+\frac{\pi}{2})|^2-|E_{o1}(\theta_1-\frac{\pi}{2})|^2]$$
$$\frac{\mathrm{d}}{\mathrm{d}\theta_2}|E_{o1}|^2=\frac{1}{2}[|E_{o1}(\theta_2+\frac{\pi}{2})|^2-|E_{o1}(\theta_2-\frac{\pi}{2})|^2]$$

The same situation applies to $|E_{o2}|^2$, and different outputs can be selected by setting different $Z$ values. For example, when $Z=\left[\begin{matrix} 1 & 0 \\ 0 & -1 \end{matrix}\right]$, $f=|E_{o1}|^2-|E_{o2}|^2$ is satisfied by the parameter-shift rule, too.

This can be extended to the case of mesh composed of MZIs, and the parameter-shift method can be used to calculate gradients for hidden layers in unitary ONNs.

\subsubsection{Complex Output of ONN Layer}
\begin{figure}[htbp]
  \centering
  \includegraphics[trim={0cm 14cm 0cm 0cm}, clip, width=0.8\textwidth]{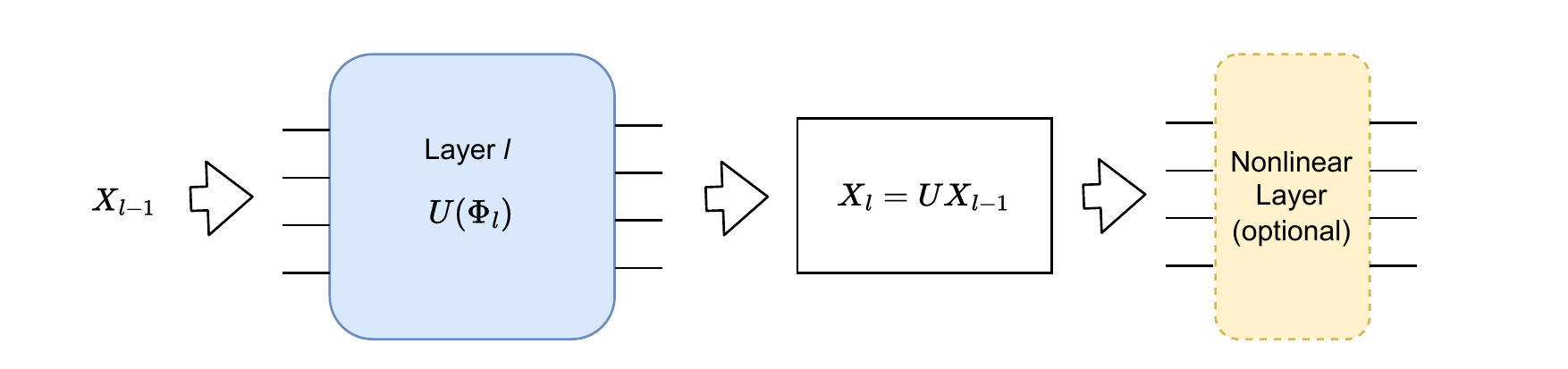}
  \caption{Schematic diagram of a optical layer with complex-valued output. All light information with phase is used as the output of the layer and input to the next layer.}
  \label{fig:3}
\end{figure}

However, in some cases like Fig. \ref{fig:3}, the output of the hidden layer in ONN is not measured but directly fed into the next layer. Under such conditions, the output of that layer is not intensity but rather a complex-valued electric field, which does not fit the calculation form of parameter-shift method. Therefore, we need to find a way to calculate the gradient of the complex output of ONN layer.

We define: 
$$f(\theta) = U_LU_{L-1}...U_j (\theta)...U_1 E$$

$$U_j(\theta) =\left[\begin{matrix} e^{i\theta} & 0 \\ 0 & 1 \end{matrix}\right] = e^{-ia\theta G}=e^{i\theta G}, where \> G=\left[\begin{matrix} 1 & 0 \\ 0 & 0 \end{matrix}\right]$$

Then the gradient is calculated to be:
$$\frac{df(\theta)}{d\theta}=U_LU_{L-1}...\frac{dU_j(\theta)}{d\theta}...U_1E$$

We have 

$$\frac{dU_j(\theta)}{d\theta}=iGU_j(\theta)$$
$$U_j(\pi/2)=I-(1-i)G$$
$$G=\frac{I-U_j (\pi/2)}{1-i}$$

The gradient can be evaluated by :

\begin{equation*} 
\begin{split}
\frac{df(\theta)}{d\theta}&=\frac{i}{1-i}U_LU_{L-1}...(I-U_j(\pi/2))U_j(\theta)...U_1E \\ 
&=\frac{i}{1-i}[f(\theta)-f(\theta+\pi/2)]
\end{split}
\end{equation*}

So for the case of complex output, the parameter-shift rule is:

$$\frac{df(\theta)}{d\theta} =\frac{1-i}{2}[f(\theta+\pi/2)-f(\theta)]$$

Through the study of the above two cases, we can use the parameter-shift method to calculate the gradients of the parameters in the unitary ONNs, thereby enabling backpropagation training for the optical neural network.

\section{Conclusion}
This paper has explored the application of the parameter-shift rule (PSR) for calculate the gradients of Unitary Optical Neural Networks (UONNs). The parameter-shift rule, a technique allowing for analytical gradient calculation by evaluating a function at shifted parameter values, was introduced as a promising alternative.We demonstrated how the PSR can be utilized to compute analytical gradients for the phase shifters within MZIs, which are fundamental components of UONNs. Crucially, we addressed two distinct scenarios: the calculation of gradients for real-valued outputs (e.g., light intensity measured by photodiodes) and for complex-valued outputs (e.g., electric field amplitudes passed between layers).

For real-valued outputs, we showed that the gradient of the output intensity with respect to a phase parameter $\theta$ can be determined by the formula $\frac{1}{2}[f(\theta+\frac{\pi}{2})-f(\theta-\frac{\pi}{2})]$, leveraging the Hermitian nature of the phase shifter's generator. For complex-valued outputs, which are essential for multi-layer ONNs where intermediate signals are not measured, we derived a modified PSR, yielding the gradient as $\frac{1-i}{2}[f(\theta+\pi/2)-f(\theta)]$. These findings confirm that the PSR provides a viable and exact method for gradient computation in UONNs, applicable to both measurable outputs and intermediate complex field values.

In conclusion, the parameter-shift rule offers a powerful and analytically precise approach for calculate the gradients of unitary optical neural networks. By enabling direct gradient calculation from hardware measurements or simulated complex outputs, it circumvents the significant implementation hurdles of all-optical backpropagation and avoids the approximation errors often associated with methods like finite differences. This work underscores the potential of leveraging mathematical frameworks like the PSR to advance the development and practical implementation of optical computing and optical artificial intelligence, paving the way for more efficient and robust training of future optical neural network architectures.







\end{document}